\documentclass{intech09}

\booktitle{Stochastic Control} 

\chaptertitle{Geometrical Derivation of Equilibrium Distributions
in some Stochastic Systems} 

\authors{Lopez-Ruiz Ricardo* and Sanudo Jaime}

\dxdoi{}

\begin{document}

\maketitle

\section{Introduction}
\label{Sec0}

Classical statistical physics deals with statistical systems in equilibrium.
The ensemble theory offers a useful framework that allows to characterize and
to work out the properties of this type of systems \cite{huang1987}.
Two fundamental distributions to describe situations in equilibrium are
the Boltzmann-Gibbs (exponential) distribution and the Maxwellian (Gaussian) distribution.
The first one represents the distribution of the energy states of a system and
the second one fits the distribution of velocities in an ideal gas.
They can be explained from different perspectives. In the physics of equilibrium, they are
usually obtained from the principle of maximum entropy \cite{jaynes1957}.
In the physics out of equilibrium,
there have recently been proposed two nonlinear models that explain the decay of any initial
distribution to these asymptotic equilibria \cite{lopez2012,shivanian2012}.

In this chapter, these distributions are alternatively obtained from a geometrical
interpretation of different multi-agent systems evolving in phase space under the
hypothesis of equiprobability. Concretely, an economic context is used to illustrate
this derivation. Thus, from a macroscopic point of view, we consider that markets
have an intrinsic stochastic ingredient as a consequence of the interaction of
an undetermined ensemble of agents that trade and perform an undetermined
number of commercial transactions at each moment. A kind of models considering this
unknowledge associated to markets are the gas-like models \cite{mantegna1999,yakovenko2009}.
These random models interpret economic exchanges of money between agents similarly
to collisions in a gas where particles share their energy.
In order to explain the two before mentioned
statistical behaviors, the Boltzmann-Gibbs and Maxwellian distributions,
we will not suppose any type of interaction between the agents. The geometrical constraints
and the hypothesis of equiprobability will be enough to explain these distributions
in a situation of statistical equilibrium.

Thus, the Boltzmann-Gibbs (exponential) distribution is derived in Section \ref{Sec1}
from the geometrical properties of the volume of an $N$-dimensional pyramid
or from the properties of the surface of an $N$-dimensional
hyperplane \cite{lopezruiz2009,lopezruiz2008}.
In both cases, the motivation will be a multi-agent economic system
with an open or closed economy, respectively.
The Maxwellian (Gaussian) distribution is derived in Section \ref{Sec2}
from geometrical arguments over the volume or
the surface of an $N$-sphere \cite{lopezruiz2009,lopezruiz2007}.
In this case, the motivation will be a multi-particle gas system in contact
with a heat reservoir (non-isolated or open system) or with a fixed energy
(isolated or closed system), respectively.
And finally, in Section \ref{Sec3},
the general equilibrium distribution for a set of many identical interacting agents
obeying a global additive constraint is also derived \cite{lopezruiz2009}.
This distribution will be related with the Gamma-like distributions found in several multi-agent
economic models. Other two geometrical collateral results, namely the formula for the volume of
high-dimensional symmetrical bodies and an alternative image of the canonical ensemble,
are proposed in Section \ref{Sec4}. And last Section embodies the conclusions.

\section{Derivation of the Boltzmann-Gibbs distribution}
\label{Sec1}

We proceed to derive here the Boltzmann-Gibbs distribution in two different
physical situations with an economic inspiration. The first one considers an
ensemble of economic agents that share a variable amount of money (open systems)
and the second one deals with the conservative case where the total wealth is fixed
(closed systems).

\subsection{Multi-agent economic open systems}

Here we assume $N$ agents, each one with coordinate $x_i$, $i=1,\ldots,N$,
with $x_i\geq 0$ representing the wealth or money of the agent $i$,
and a total available amount of money $E$:
\begin{equation}
x_1+x_2+\cdots +x_{N-1}+x_N \leq E.
\label{eq-e}
\end{equation}
Under random or deterministic evolution rules for the exchanging of money among agents,
let us suppose that this system evolves in the interior of the $N$-dimensional pyramid
given by Eq. (\ref{eq-e}). The role of a heat reservoir, that in this model
supplies money instead of energy, could be played by the state or by the bank system
in western societies.
The formula for the volume $V_N(E)$ of an equilateral $N$-dimensional pyramid
formed by $N+1$ vertices linked by $N$ perpendicular sides of length $E$ is
\begin{equation}
V_N(E) = {E^N\over N!}.
\label{eq-S_n1}
\end{equation}
We suppose that each point on the $N$-dimensional pyramid is equiprobable,
then the probability $f(x_i)dx_i$ of finding
the agent $i$ with money $x_i$ is proportional to the
volume formed by all the points into the $(N-1)$-dimensional pyramid
having the $i$th-coordinate equal to $x_i$.
We show now that $f(x_i)$ is the Boltzmann factor
(or the Maxwell-Boltzmann distribution), with the normalization condition
\begin{equation}
\int_{0}^Ef(x_i)dx_i = 1.
\label{eq-p_n1}
\end{equation}

If the $i$th agent has coordinate $x_i$, the $N-1$ remaining agents
share, at most, the money $E-x_i$ on the $(N-1)$-dimensional pyramid
\begin{equation}
x_1+x_2 \cdots +x_{i-1} + x_{i+1} \cdots +x_N\leq E-x_i,
\label{eq-e1}
\end{equation}
whose volume is $V_{N-1}(E-x_i)$.
It can be easily proved that
\begin{equation}
V_N(E) = \!\int_{0}^{E}\!V_{N-1}(E-x_i) {dx_i }.
\label{eq-theta11}
\end{equation}

Hence, the volume of the $N$-dimensional pyramid for which the $i$th
coordinate is between $x_i$ and $x_i+dx_i$ is $V_{N-1}(E-x_i)dx_i$.
We normalize it to satisfy Eq.~(\ref{eq-p_n1}), and obtain
\begin{equation}
f(x_i) = {V_{N-1}(E-x_i)\over V_N(E)},
\label{eq-f_n1}
\end{equation}
whose final form, after some calculation is
\begin{equation}
f(x_i) = NE^{-1}\Big(1-{x_i\over E} \Big)^{N-1},
\label{eq-mm1}
\end{equation}
If we call $\epsilon$ the mean wealth per agent,
$E=N\epsilon$, then in the limit of large $N$
we have
\begin{equation}
\lim_{N\gg 1}\left(1-{x_i\over E}\right)^{N-1}
\simeq e^{-{x_i/\epsilon}}.
\label{eq-ee1}
\end{equation}
The Boltzmann factor $e^{-{x_i/\epsilon}}$ is found
when $N\gg 1$ but, even for small $N$, it can be a good approximation
for agents with low wealth. After substituting Eq.~(\ref{eq-ee1})
into Eq.~(\ref{eq-mm1}), we obtain the Maxwell-Boltzmann distribution
in the asymptotic regime $N\rightarrow\infty$ (which also implies $E\rightarrow\infty$):
\begin{equation}
f(x)dx = {1\over \epsilon}\,e^{-{x/\epsilon}}dx,
\label{eq-gauss11}
\end{equation}
where the index $i$ has been removed because the distribution is the same for each agent,
and thus the wealth distribution can be obtained by averaging over all the agents.
This distribution has been found to fit the real distribution of incomes
in western societies \cite{dragulescu2000,chakraborti2000}.

This geometrical image of the volume-based statistical ensemble \cite{lopezruiz2009}
allows us to recover the same result than that obtained
from the microcanonical ensemble \cite{lopezruiz2008} that we show in the next section.

\subsection{Multi-agent economic closed systems}

Here, we derive the Boltzmann-Gibbs distribution by considering the system
in isolation, that is, a closed economy. Without loss of generality,
let us assume $N$ interacting economic agents, each one with coordinate
$x_i$, $i=1,\ldots,N$, with $x_i\geq 0$, and where $x_i$ represents an amount of money.
If we suppose now that the total amount of money $E$ is conserved,
\begin{equation}
x_1+x_2+\cdots +x_{N-1}+x_N = E,
\label{eq-E}
\end{equation}
then this isolated system
evolves on the positive part of an equilateral $N$-hyperplane.
The surface area $S_N(E)$ of an equilateral
$N$-hyperplane of side $E$ is given by
\begin{equation}
S_N(E) = {\sqrt{N}\over (N-1)!}\;E^{N-1}.
\label{eq-S_n2}
\end{equation}
Different rules, deterministic or random, for the exchange of money between agents
can be given. Depending on these rules, the system can visit the
$N$-hyperplane in an equiprobable manner or not.
If the ergodic hypothesis is
assumed, each point on the $N$-hyperplane is equiprobable.
Then the probability $f(x_i)dx_i$ of finding
agent $i$ with money $x_i$ is proportional to the
surface area formed by all the points on the $N$-hyperplane having the $i$th-coordinate
equal to $x_i$. We show that $f(x_i)$ is the Boltzmann-Gibbs distribution (the Boltzmann factor),
with the normalization condition (\ref{eq-p_n1}).

If the $i$th agent has coordinate $x_i$, the $N-1$ remaining agents
share the money $E-x_i$ on the $(N-1)$-hyperplane
\begin{equation}
x_1+x_2 \cdots +x_{i-1} + x_{i+1} \cdots +x_N= E-x_i,
\label{eq-e11}
\end{equation}
whose surface area is $S_{N-1}(E-x_i)$.
If we define the coordinate $\theta_N$ as satisfying
\begin{equation}
\sin\theta_N = \sqrt{N-1 \over N},
\label{eq-theta}
\end{equation}
it can be easily shown that
\begin{equation}
S_N(E) = \!\int_{0}^{E}\!S_{N-1}(E-x_i) {dx_i \over \sin\theta_N}.
\label{eq-theta122}
\end{equation}
Hence, the surface area of the $N$-hyperplane for which the $i$th coordinate is
between $x_i$ and $x_i+dx_i$ is proportional to $S_{N-1}(E-x_i)dx_i/\sin\theta_N$.
If we take into account the normalization condition (\ref{eq-p_n1}), we obtain
\begin{equation}
f(x_i) = {1\over S_N(E)}
{S_{N-1}(E-x_i)\over \sin\theta_N},
\label{eq-f_n}
\end{equation}
whose form after some calculation is
\begin{equation}
f(x_i) = (N-1)E^{-1}\Big(1-{x_i\over E} \Big)^{N-2},
\label{eq-mmm}
\end{equation}
If we call $\epsilon$ the mean wealth per agent,
$E=N\epsilon$, then in the limit of large $N$
we have
\begin{equation}
\lim_{N\gg 1}\left(1-{x_i\over E}\right)^{N-2}
\simeq e^{-{x_i/\epsilon}}.
\label{eq-eee}
\end{equation}
As in the former section, the Boltzmann factor $e^{-{x_i/\epsilon}}$ is found
when $N\gg 1$ but, even for small $N$, it can be a good approximation
for agents with low wealth. After substituting Eq.~(\ref{eq-e})
into Eq.~(\ref{eq-mmm}), we obtain the Boltzmann distribution (\ref{eq-gauss11})
in the limit $N\rightarrow\infty$ (which also implies $E\rightarrow\infty$).
This asymptotic result reproduces the distribution of real economic
data \cite{dragulescu2000} and also the results obtained in several models
of economic agents with deterministic, random or chaotic exchange
interactions \cite{yakovenko2009,gonzalez2008,pellicer2010}.

Depending on the physical situation, the mean wealth per agent $\epsilon$
takes different expressions and interpretations.
For instance,
we can calculate the dependence of $\epsilon$ on the temperature, which
in the microcanonical ensemble is defined by the derivative of the entropy
with respect to the energy. The entropy can be written as $S=-kN\int_{0}
^{\infty} f(x)\ln f(x)\,dx$, where $f(x)$ is given by Eq.~(\ref{eq-gauss11})
and $k$ is Boltzmann's constant.
If we recall that $\epsilon=E/N$, we obtain
\begin{equation}
S(E)= kN\ln {E\over N} + kN.
\end{equation}
The calculation of the temperature $T$ gives
\begin{equation}
T^{-1}= \left({\partial S\over \partial E} \right)_N = {kN\over E} = {k\over \epsilon}.
\end{equation}
Thus $\epsilon=kT$, and the Boltzmann-Gibbs distribution
is obtained in its usual form:
\begin{equation}
f(x)dx = {1\over kT}\,e^{-x/kT}dx.
\end{equation}

\section{Derivation of the Maxwellian distribution}
\label{Sec2}

We proceed to derive here the Maxwellian distribution in two different
physical situations with inspiration in the theory of ideal gases.
The first one considers an ideal gas with a variable energy (open systems)
and the second one deals with the case of a gas with a fixed energy (closed systems).

\subsection{Multi-particle open systems}

Let us suppose a one-dimensional ideal gas of $N$ non-identical
classical particles with masses $m_i$, with $i=1,\ldots,N$, and total
maximum energy $E$. If particle
$i$ has a momentum $m_iv_i$, we define a kinetic energy:
\begin{equation}
K_i \equiv p_i^2 \equiv {1 \over 2}{ m_iv_i^2},
\label{eq-p_i}
\end{equation}
where $p_i$ is the square root of the kinetic energy $K_i$.
If the total maximum energy is defined as $E \equiv R^2$, we have
\begin{equation}
p_1^2+p_2^2+\cdots +p_{N-1}^2+p_N^2 \leq R^2.
\label{eq-Ee}
\end{equation}
We see that the system has accessible states with different energy, which can be
supplied by a heat reservoir. These states are all those enclosed into the volume
of the $N$-sphere given by Eq. (\ref{eq-Ee}).
The formula for the volume $V_N(R)$
of an $N$-sphere of radius $R$ is
\begin{equation}
V_N(R) = {\pi^{N\over 2}\over \Gamma({N\over 2}+1)}R^{N},
\label{eq-S_n3}
\end{equation}
where $\Gamma(\cdot)$ is the gamma function. If we suppose that each point
into the $N$-sphere is equiprobable, then the probability $f(p_i)dp_i$ of finding
the particle $i$ with coordinate $p_i$ (energy $p_i^2$) is proportional to the
volume formed by all the points on the $N$-sphere having the $i$th-coordinate
equal to $p_i$.
We proceed to show that $f(p_i)$ is the Maxwellian
distribution, with the normalization condition
\begin{equation}
\int_{-R}^Rf(p_i)dp_i = 1.
\label{eq-p_n2}
\end{equation}
If the $i$th particle has coordinate $p_i$, the $(N-1)$ remaining particles
share an energy less than the maximum energy $R^2-p_i^2$ on the $(N-1)$-sphere
\begin{equation}
p_1^2+p_2^2 \cdots +p_{i-1}^2 + p_{i+1}^2 \cdots +p_N^2 \leq R^2-p_i^2,
\label{eq-E12}
\end{equation}
whose volume is $V_{N-1}(\sqrt{R^2-p_i^2})$.
It can be easily proved that
\begin{equation}
V_N(R) = \!\int_{-R}^{R}\!V_{N-1}(\sqrt{R^2-p_i^2})dp_i.
\label{eq-theta13}
\end{equation}
Hence, the volume of the $N$-sphere for which the $i$th coordinate is
between $p_i$ and $p_i+dp_i$ is $V_{N-1}(\sqrt{R^2-p_i^2})dp_i$.
We normalize it to satisfy Eq.~(\ref{eq-p_n2}), and obtain
\begin{equation}
f(p_i) = {V_{N-1}(\sqrt{R^2-p_i^2})\over V_N(R)},
\label{eq-f_n2}
\end{equation}
whose final form, after some calculation is
\begin{equation}
f(p_i) = C_N R^{-1}\Big(1-{p_i^2\over R^2} \Big)^{N-1\over 2},
\label{eq-mmm1}
\end{equation}
with
\begin{equation}
C_N = {1\over\sqrt{\pi}}{\Gamma({N+2\over 2})\over \Gamma({N+1\over 2})}.
\label{eq-cnn}
\end{equation}
For $N\gg 1$, Stirling's approximation can be applied to
Eq.~(\ref{eq-cnn}), leading to
\begin{equation}
\lim_{N\gg 1} C_N \simeq {1\over\sqrt{\pi}}\sqrt{N\over 2}.
\label{eq-ccc}
\end{equation}
If we call $\epsilon$ the mean energy per particle,
$E=R^2=N\epsilon$, then in the limit of large $N$ we have
\begin{equation}
\lim_{N\gg 1}\left(1-{p_i^2\over R^2}\right)^{N-1\over 2}
\simeq e^{-{p_i^2/2\epsilon}}.
\label{eq-eee1}
\end{equation}
The factor $e^{-{p_i^2/2\epsilon}}$ is found
when $N\gg 1$ but, even for small $N$, it can be a good approximation
for particles with low energies.
After substituting Eqs.~(\ref{eq-ccc})--(\ref{eq-eee1})
into Eq.~(\ref{eq-mmm1}), we obtain the Maxwellian distribution in the asymptotic regime $N\rightarrow\infty$
(which also implies $E\rightarrow\infty$):
\begin{equation}
f(p)dp = \sqrt{1\over 2\pi\epsilon}\,e^{-{p^2/2\epsilon}}dp,
\label{eq-gauss}
\end{equation}
where the index $i$ has been removed because the distribution is the same for each particle,
and thus the velocity distribution can be obtained by averaging
over all the particles.

This newly shows that the geometrical image of the volume-based statistical ensemble \cite{lopezruiz2009}
allows us to recover the same result than that obtained
from the microcanonical ensemble \cite{lopezruiz2007} that it is presented in the next section.

\subsection{Multi-particle closed systems}

We start by assuming a one-dimensional ideal gas of $N$ non-identical
classical particles with masses $m_i$, with $i=1,\ldots,N$, and total
energy $E$. If particle $i$ has a momentum $m_iv_i$, newly we define a
kinetic energy $K_i$ given by Eq. (\ref{eq-p_i}), where $p_i$ is the square
root of $K_i$. If the total energy is defined as $E \equiv R^2$, we have
\begin{equation}
p_1^2+p_2^2+\cdots +p_{N-1}^2+p_N^2 = R^2.
\label{eq-E1}
\end{equation}
We see that the isolated system evolves on the surface of an $N$-sphere.
The formula for the surface area $S_N(R)$
of an $N$-sphere of radius $R$ is
\begin{equation}
S_N(R) = {2\pi^{N\over 2}\over \Gamma({N\over 2})}R^{N-1},
\label{eq-S_n4}
\end{equation}
where $\Gamma(\cdot)$ is the gamma function. If the ergodic hypothesis is
assumed, that is, each point on the $N$-sphere is equiprobable,
then the probability $f(p_i)dp_i$ of finding
the particle $i$ with coordinate $p_i$ (energy $p_i^2$) is proportional to the
surface area formed by all the points on the $N$-sphere having the $i$th-coordinate
equal to $p_i$.
Our objective is to show that $f(p_i)$ is the Maxwellian
distribution, with the normalization condition (\ref{eq-p_n2}).

If the $i$th particle has coordinate $p_i$, the $(N-1)$ remaining particles
share the energy $R^2-p_i^2$ on the $(N-1)$-sphere
\begin{equation}
p_1^2+p_2^2 \cdots +p_{i-1}^2 + p_{i+1}^2 \cdots +p_N^2= R^2-p_i^2,
\label{eq-E11}
\end{equation}
whose surface area is $S_{N-1}(\sqrt{R^2-p_i^2})$.
If we define the coordinate $\theta$ as satisfying
\begin{equation}
R^2\cos^2\theta = R^2-p_i^2,
\label{eq-theta12}
\end{equation}
then
\begin{equation}
Rd\theta = {dp_i \over (1-{p_i^2\over R^2})^{1/2}}.
\label{eq-diftheta}
\end{equation}
It can be easily proved that
\begin{equation}
S_N(R) = \!\int_{-\pi/2}^{\pi/2}\!S_{N-1}(R\cos\theta)R d\theta.
\label{eq-theta14}
\end{equation}
Hence, the surface area of the $N$-sphere for which the $i$th coordinate is
between $p_i$ and $p_i+dp_i$ is $S_{N-1}(R\cos\theta)Rd\theta$.
We rewrite the surface area
as a function of $p_i$,
normalize it to satisfy Eq.~(\ref{eq-p_n2}), and obtain
\begin{equation}
f(p_i) = {1\over S_N(R)}
{S_{N-1}(\sqrt{R^2-p_i^2})\over (1-{p_i^2\over R^2})^{1/2}},
\label{eq-f_n3}
\end{equation}
whose final form, after some calculation is
\begin{equation}
f(p_i) = C_N R^{-1}\Big(1-{p_i^2\over R^2} \Big)^{N-3\over 2},
\label{eq-mm}
\end{equation}
with
\begin{equation}
C_N = {1\over\sqrt{\pi}}{\Gamma({N\over 2})\over \Gamma({N-1\over 2})}.
\label{eq-cn}
\end{equation}
For $N\gg 1$, Stirling's approximation can be applied to
Eq.~(\ref{eq-cn}), leading to
\begin{equation}
\lim_{N\gg 1} C_N \simeq {1\over\sqrt{\pi}}\sqrt{N\over 2}.
\label{eq-cc}
\end{equation}
If we call $\epsilon$ the mean energy per particle,
$E=R^2=N\epsilon$, then in the limit of large $N$
we have
\begin{equation}
\lim_{N\gg 1}\left(1-{p_i^2\over R^2}\right)^{N-3\over 2}
\simeq e^{-{p_i^2/2\epsilon}}.
\label{eq-ee}
\end{equation}
As in the former section, the Boltzmann factor $e^{-{p_i^2/2\epsilon}}$ is found
when $N\gg 1$ but, even for small $N$, it can be a good approximation
for particles with low energies.
After substituting Eqs.~(\ref{eq-cc})--(\ref{eq-ee})
into Eq.~(\ref{eq-mm}), we obtain the Maxwellian distribution (\ref{eq-gauss})
in the asymptotic regime $N\rightarrow\infty$
(which also implies $E\rightarrow\infty$).

Depending on the physical situation the mean energy per particle $\epsilon$
takes different expressions. For an isolated one-dimensional gas
we can calculate the dependence of $\epsilon$ on the temperature, which
in the microcanonical ensemble is defined by differentiating the entropy
with respect to the energy. The entropy can be written as $S=-kN\!\int_{-\infty}
^{\infty} f(p)\ln f(p)\,dp$, where $f(p)$ is given by Eq.~(\ref{eq-gauss})
and $k$ is the Boltzmann constant.
If we recall that $\epsilon=E/N$, we obtain
\begin{equation}
S(E)= {1\over 2}kN\ln\left({E\over N} \right) + {1\over 2}kN(\ln(2\pi)-1).
\end{equation}
The calculation of the temperature $T$ gives
\begin{equation}
T^{-1}= \left({\partial S\over \partial E} \right)_N = {kN\over 2E} = {k\over 2\epsilon}.
\end{equation}
Thus $\epsilon=kT/2$, consistent with the equipartition theorem.
If $p^2$ is replaced by ${1\over 2}mv^2$, the Maxwellian
distribution is a function of particle velocity, as it is usually given
in the literature:
\begin{equation}
g(v)dv = \sqrt{m\over 2\pi kT}\,e^{-{mv^2/2kT}}dv.
\end{equation}

\section{General derivation of the equilibrium distribution}
\label{Sec3}

In this section, we are interested in the same problem above presented but in a general way.
We address this question in the volume-based statistical framework.

Let $b$ be a positive real constant (cases $b=1,2$ have been indicated in the former sections).
If we have a set of positive variables $(x_1,x_2,\ldots,x_N)$ verifying the constraint
\begin{equation}
x_1^b+x_2^b+\cdots +x_{N-1}^b+x_N^b \leq E
\label{eq-Ek}
\end{equation}
with an adequate mechanism assuring
the equiprobability of all the possible states $(x_1,x_2,\ldots,x_N)$
into the volume given by expression (\ref{eq-Ek}),
will we have for the generic variable $x$ the distribution
\begin{equation}
f(x)dx \sim \epsilon^{-1/b}\,e^{-{x^b/b\epsilon}}dx,
\label{eq-gaussn}
\end{equation}
when we average over the ensemble in the limit $N,E\rightarrow\infty$,
with $E=N\epsilon$, and constant $\epsilon$?.
Now it is shown that the answer is affirmative.
Similarly, we claim that if the weak inequality (\ref{eq-Ek}) is transformed in
equality the result will be the same, as it has been proved for the cases $b=1,2$
in Refs. \cite{lopezruiz2008,lopezruiz2007}.

From the cases $b=1,2$, (see Eqs. (\ref{eq-f_n1}) and (\ref{eq-f_n2})),
we can extrapolate the general formula that will give
us the statistical behavior $f(x)$ of the generic variable $x$,
when the system runs equiprobably into the volume defined by a constraint of type (\ref{eq-Ek}).
The probability $f(x)dx$ of finding
an agent with generic coordinate $x$ is proportional to the
volume $V_{N-1}((E-x^b)^{1/b})$ formed by all the points into the $(N-1)$-dimensional
symmetrical body limited by the constraint $(E-x^b)$.
Thus, the $N$-dimensional volume can be written as
\begin{equation}
V_N(E^{1/b})= \int_{0}^{E^{1/b}}\,V_{N-1}((E-x^b)^{1/b})\,dx.
\label{eq-volumen}
\end{equation}
Taking into account the normalization condition $\int_{0}^{E^{1/b}}f(x)dx = 1$,
the expression for $f(x)$ is obtained:
\begin{equation}
f(x) = {V_{N-1}((E-x^b)^{1/b})\over V_N(E^{1/b})}.
\label{eq-volume3}
\end{equation}
The $N$-dimensional volume, $V_N(b,\rho)$, of a $b$-symmetrical body with side of length $\rho$ is
proportional to the term $\rho^N$ and to a coefficient $g_b(N)$ that depends on $N$:
\begin{equation}
V_N(b,\rho)=g_b(N)\,\rho^N.
\label{eq-volumenn}
\end{equation}
The parameter $b$ indicates the original equation (\ref{eq-Ek}) that defines the
boundaries of the volume $V_N(b,\rho)$. Thus, for instance, from Eq. (\ref{eq-S_n1}),
we have $g_{b=1}(N) =  1/ N!$.

Coming back to Eq. (\ref{eq-volume3}), we can manipulate $V_{N}((E-x^b)^{1/b})$
to obtain (the index $b$ is omitted in the formule of $V_N$):
\begin{equation}
V_N((E-x^b)^{1/b})=g_b(N)\,\left[(E-x^b)^{1/b}\right]^N=g_b(N)\,E^{N\over b}\,
\left(1-{x^b\over E}\right)^{N\over b}.
\label{eq-volumexb}
\end{equation}
If we suppose $E=N\epsilon$, then $\epsilon$ represents the mean value of $x^b$
in the collectivity, that is, $\epsilon=<x^b>$. If $N$ tends toward infinity,
it results:
\begin{equation}
\lim_{N\gg 1}\left(1-{x^b\over E}\right)^{N\over b} \, =\, e^{-x^b/ b\epsilon}.
\label{eq-volumexb1}
\end{equation}
Thus,
\begin{equation}
V_N((E-x^b)^{1/b})=V_N(E^{1/b})\,e^{-x^b/ b\epsilon}.
\label{eq-volumexb2}
\end{equation}
Substituting this last expression in formula (\ref{eq-volume3}),
the exact form for $f(x)$ is found in the thermodynamic limit
($N,E\rightarrow\infty$):
\begin{equation}
f(x)dx = c_b \,\epsilon^{-1/b}\,e^{-{x^b/b\epsilon}}dx,
\label{eq-fx}
\end{equation}
with $c_b$ given by
\begin{equation}
c_b\,=\,{g_b(N-1)\over g_b(N)\,N^{1/b}}.
\label{eq-cb}
\end{equation}
Hence, the conjecture (\ref{eq-gaussn}) is proved.

Doing a thermodinamical simile,
we can calculate the dependence of $\epsilon$ on the temperature
by differentiating the entropy with respect to the energy.
The entropy can be written as $S=-kN\!\int_{0}
^{\infty} f(x)\ln f(x)\,dx$, where $f(x)$ is given by Eq.~(\ref{eq-fx})
and $k$ is the Boltzmann constant.
If we recall that $\epsilon=E/N$, we obtain
\begin{equation}
S(E)= {kN\over b}\ln\left({E\over N} \right) + {kN\over b}(1-b\ln c_b),
\end{equation}
where it has been used that $\epsilon=<x^b>=\!\int_{0} ^{\infty} x^bf(x)dx$.

The calculation of the temperature $T$ gives
\begin{equation}
T^{-1}= \left({\partial S\over \partial E} \right)_N = {kN\over bE} = {k\over b\epsilon}.
\end{equation}
Thus $\epsilon=kT/b$, a result that recovers the theorem of equipartition of energy
for the quadratic case $b=2$.
The distribution for all $b$ is finally obtained:
\begin{equation}
f(x)dx = c_b\left({b\over kT}\right)^{1/b}\,e^{-x^b/kT}dx.
\end{equation}

\subsection{General relationship between geometry and economic gas models}

If we perform the change of variables $y=\epsilon^{-1/b}x$ in the normalization
condition of $f(x)$, $\!\int_{0} ^{\infty} f(x)dx=1$, where $f(x)$ is expressed in (\ref{eq-fx}),
we find that
\begin{equation}
c_b=\left[\!\int_{0} ^{\infty} e^{-y^b/b}\,dy\right]^{-1}.
\label{eq-cb1}
\end{equation}
If we introduce the new variable $z=y^b/b$, the distribution $f(x)$ as function of $z$ reads:
\begin{equation}
f(z)dz = {c_b\over b^{1-{1\over b}}} \,z^{{1\over b}-1}\,e^{-z}\,dz.
\label{eq-fz}
\end{equation}
Let us observe that the Gamma function appears in the normalization condition,
\begin{equation}
\int_{0} ^{\infty} f(z)dz={c_b\over b^{1-{1\over b}}} \,\int_{0} ^{\infty}
\,z^{{1\over b}-1}\,e^{-z}\,dz = {c_b\over b^{1-{1\over b}}}\,\Gamma\left({1\over b}\right)=1.
\label{eq-fz1}
\end{equation}
This implies that
\begin{equation}
c_b={b^{1-{1\over b}} \over \Gamma\left({1\over b}\right)}.
\label{eq-cb2}
\end{equation}
By using Mathematica the positive constant $c_b$ is plotted versus $b$ in Fig. 1.
We see that $\lim_{b\rightarrow 0}c_b=\infty$, and that $\lim_{b\rightarrow \infty}c_b=1$.
The minimum of $c_b$ is reached for $b=3.1605$, taking the value $c_b=0.7762$.
Still further, we can calculate from Eq. (\ref{eq-cb2})
the asymptotic dependence of $c_b$ on b:
\begin{eqnarray}
\lim_{b\rightarrow 0}c_b & = & \sqrt{1\over 2\pi}\,\sqrt{b}\,e^{1/b}\left(1-{b\over 12}+
\cdots\right), \label{eq-cb3}\\
&& \nonumber\\
\lim_{b\rightarrow \infty}c_b & = & b^{-1/b}\left(1 + {\gamma\over b}+\cdots\right),\label{eq-cb4}
\end{eqnarray}
where $\gamma$ is the Euler constant, $\gamma=0.5772$. The asymptotic function (\ref{eq-cb3}) is
obtained after substituting in (\ref{eq-cb2}) the value of $\Gamma(1/b)$ by $(1/b-1)!$, and performing the
Stirling approximation on this last expression, knowing that $1/b\rightarrow\infty$. The function
(\ref{eq-cb4}) is found after looking for the first Taylor expansion terms of the Gamma function
around the origin $x=0$. They can be derived from the Euler's reflection formula,
$\Gamma(x)\Gamma(1-x)=\pi/\sin(\pi x)$. We obtain $\Gamma(x\rightarrow 0)=x^{-1}+\Gamma'(1)+\cdots$.
From here, recalling that $\Gamma'(1)=-\gamma$, we get $\Gamma(1/b)=b-\gamma+\cdots$,
when $b\rightarrow\infty$. Although this last term of the Taylor expansion, $-\gamma$,
is negligible we maintain it in expression (\ref{eq-cb4}). The only minimum of $c_b$ is reached for
the solution $b=3.1605$ of the equation $\psi(1/b)+\log b+b-1=0$, where $\psi(\cdot)$ is the
digamma function (see Fig. 1).

Let us now recall two interesting statistical economic models that display a statistical behavior
given by distributions nearly to the form (\ref{eq-fz}), that is, the standard Gamma distributions
with shape parameter $1/b$,
\begin{equation}
f(z)dz = {1\over \Gamma({1\over b})} \,z^{{1\over b}-1}\,e^{-z}\,dz.
\label{eq-fg}
\end{equation}

\begin{figure}[t]
\caption{Normalization constant $c_b$ versus $b$, calculated from Eq. (\ref{eq-cb2}).
The asymptotic behavior is: $\lim_{b\rightarrow 0}c_b=\infty$, and $\lim_{b\rightarrow \infty}c_b=1$.
This last asymptote is represented by the dotted line.
The minimum of $c_b$ is reached for $b=3.1605$, taking the value $c_b=0.7762$.}
\centerline{\includegraphics[width=9cm]{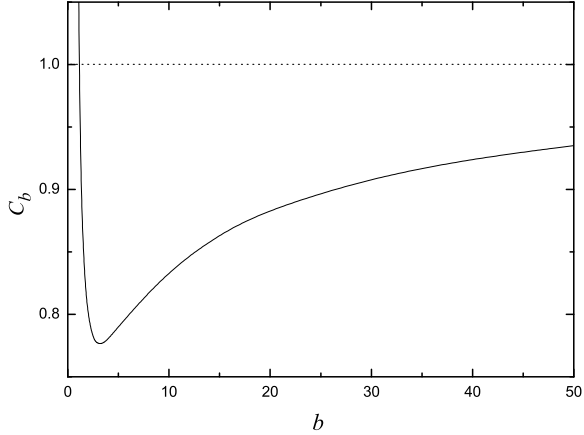}}
\label{fig1}
\end{figure}

{\bf ECONOMIC MODEL A:} The first one is the saving propensity model introduced by
Chakraborti and Chakrabarti \cite{chakraborti2000}. In this model a set of $N$ economic
agents, having each agent $i$ (with $i=1,2,\cdots,N$) an amount of money, $u_i$,
exchanges it under random binary $(i,j)$ interactions, $(u_i,u_j)\rightarrow (u_i',u_j')$,
by the following the exchange rule:
\begin{eqnarray}
u'_i & = & \lambda u_i+\epsilon(1-\lambda)(u_i+u_j), \\
u'_j & = & \lambda u_j+\bar\epsilon(1-\lambda)(u_i+u_j),
\end{eqnarray}
with $\bar\epsilon=(1-\epsilon)$, and $\epsilon$ a random number in the interval $(0,1)$.
The parameter $\lambda$, with $0<\lambda<1$, is fixed, and represents the fraction of money
saved before carrying out the transaction. Let us observe that money is conserved, i.e.,
$u_i+u_j=u_i'+u_j'$, hence in this model the economy is closed. Defining the parameter
$n(\lambda)$ as
\begin{equation}
n(\lambda)={1+2\lambda \over 1-\lambda},
\label{eq-nl1}
\end{equation}
and scaling the wealth of the agents as $\bar z=nu/<u>$, with $<u>$ representing the average money
over the ensemble of agents, it is found that the asymptotic wealth distribution in this system
is nearly obeying the standard Gamma distribution \cite{patriarca2004,calbet2011}
\begin{equation}
f(\bar z)d\bar z = {1\over \Gamma(n)} \,\bar z^{n-1}\,e^{-\bar z}\,d\bar z.
\label{eq-fg1}
\end{equation}
The case $n=1$, which means a null saving propensity, $\lambda=0$, recovers the model
of Dragulescu and Yakovenko \cite{dragulescu2000} in which the Gibbs distribution is observed.
If  we compare Eqs. (\ref{eq-fg1}) and (\ref{eq-fg}), a close relationship between this
economic model and the geometrical problem solved in the former section can be established.
It is enough to make
\begin{eqnarray}
n & = & 1/b, \label{eq-nb}\\
\bar z & = & z\label{eq-nz},
\end{eqnarray}
to have two equivalent systems. This means that, from Eq. (\ref{eq-nb}),
we can calculate $b$ from the saving parameter $\lambda$ with the formula
\begin{equation}
b={1-\lambda \over 1+2\lambda}.
\label{eq-nl11}
\end{equation}
As $\lambda$ takes its values in the interval $(0,1)$,
then the parameter $b$ also runs in the same interval $(0,1)$.
On the other hand, recalling that $z=x^b/b\epsilon$,
we can get the equivalent variable $x$ from Eq. (\ref{eq-nz}),
\begin{equation}
x=\left[{\epsilon\over <u>}\;u\;\right]^{1/b},
\label{eq-nl12}
\end{equation}
where $\epsilon$ is a free parameter that determines the mean value
of $x^b$ in the equivalent geometrical system.
Formula (\ref{eq-nl12}) means to perform the change of variables
$u_i\rightarrow x_i$, with $i=1,2,\cdots,N$, for all the particles/agents of the ensemble.
Then, we conjecture that the economic system represented by the generic
pair $(\lambda,u)$, when it is transformed in
the geometrical system given by the generic pair $(b,x)$,
as indicated by the rules (\ref{eq-nl11}) and (\ref{eq-nl12}),
runs in an equiprobable form on the surface defined by the
relationship (\ref{eq-Ek}), where the inequality has been transformed
in equality. This last detail is due to the fact the economic system is closed,
and then it conserves the total money, whose equivalent quantity in the geometrical
problem is $E$. If the economic system were open, with an upper limit in the wealth,
then the transformed system would evolve in an equiprobable way over the volume
defined by the inequality (\ref{eq-Ek}), although its statistical behavior
would continue to be the same  as it has been proved for the cases $b=1,2$
in Refs. \cite{lopezruiz2008,lopezruiz2007}.

{\bf ECONOMIC MODEL B:}
The second one is a model introduced in \cite{patriarca2006}.
In this model a set of $N$ economic
agents, having each agent $i$ (with $i=1,2,\cdots,N$) an amount of money, $u_i$,
exchanges it under random binary $(i,j)$ interactions, $(u_i,u_j)\rightarrow (u_i',u_j')$,
by the following the exchange rule:
\begin{eqnarray}
u'_i & = & u_i-\Delta u, \\
u'_j & = & u_j+\Delta u,
\end{eqnarray}
where
\begin{equation}
\Delta u=\eta(x_i-x_j)\,\epsilon\omega x_i-[1-\eta(x_i-x_j)]\,\epsilon\omega x_j,
\label{eq-delta}
\end{equation}
with $\epsilon$ a continuous uniform random number in the interval $(0,1)$.
When this variable is transformed in a Bernouilli variable, i.e. a discrete uniform random
variable taking on the values $0$ or $1$, we have the model studied by Angle \cite{angle2006},
that gives very different asymptotic results.
The exchange parameter, $\omega$,
represents the maximum fraction of wealth lost by one of the two interacting agents ($0<\omega< 1$).
Whether the agent who is going to loose part of the money is the $i$-th or the $j$-th agent,
depends nonlinearly on $(x_i-x_j)$, and this is decided by the random dichotomous function $\eta(t)$:
$\eta(t>0)=1$ (with additional probability $1/2$) and $\eta(t<0)=0$ (with additional probability $1/2$).
Hence, when $x_i>x_j$,  the value $\eta=1$ produces a wealth transfer from agent $i$ to agent $j$
with probability $1/2$, and when $x_i<x_j$,  the value $\eta=0$ produces a wealth transfer from
agent $j$ to agent $i$ with probability $1/2$.
Defining in this case the parameter $n(\omega)$ as
\begin{equation}
n(\omega)={3-2\omega \over 2\omega},
\label{eq-na1}
\end{equation}
and scaling the wealth of the agents as $\bar z=nu/<u>$, with $<u>$ representing the average money
over the ensemble of agents, it is found that the asymptotic wealth distribution in this system
is nearly to fit the standard Gamma distribution \cite{calbet2011,patriarca2006}
\begin{equation}
f(\bar z)d\bar z = {1\over \Gamma(n)} \,\bar z^{n-1}\,e^{-\bar z}\,d\bar z.
\label{eq-fg11}
\end{equation}
The case $n=1$, which means an exchange parameter $\omega=3/4$, recovers the model
of Dragulescu and Yakovenko \cite{dragulescu2000} in which the Gibbs distribution is observed.
If  we compare Eqs. (\ref{eq-fg11}) and (\ref{eq-fg}), a close relationship between this
economic model and the geometrical problem solved in the last section can be established.
It is enough to make
\begin{eqnarray}
n & = & 1/b, \label{eq-nb1}\\
\bar z & = & z\label{eq-nz1},
\end{eqnarray}
to have two equivalent systems. This means that, from Eq. (\ref{eq-nb1}),
we can calculate $b$ from the exchange parameter $\omega$ with the formula
\begin{equation}
b={2\omega \over 3-2\omega}.
\label{eq-nl22}
\end{equation}
As $\omega$ takes its values in the interval $(0,1)$,
then the parameter $b$ runs in the interval $(0,2)$.
It is curious to observe that in this model the interval $\omega\in(3/4,1)$ maps on $b\in(1,2)$,
a fact that does not occur in MODEL A.
On the other hand, recalling that $z=x^b/b\epsilon$,
we can get the equivalent variable $x$ from Eq. (\ref{eq-nz1}),
\begin{equation}
x=\left[{\epsilon\over <u>}\;u\;\right]^{1/b}.
\label{eq-nl23}
\end{equation}
where $\epsilon$ is a free parameter that determines the mean value
of $x^b$ in the equivalent geometrical system.
Formula (\ref{eq-nl23}) means to perform the change of variables
$u_i\rightarrow x_i$, with $i=1,2,\cdots,N$, for all the particles/agents of the ensemble.
Then, also in this case, we conjecture that the economic system represented by the generic
pair $(\lambda,u)$, when it is transformed in
the geometrical system given by the generic pair $(b,x)$,
as indicated by the rules (\ref{eq-nl22}) and (\ref{eq-nl23}),
runs in an equiprobable form on the surface defined by the
relationship (\ref{eq-Ek}), where the inequality has been transformed
in equality. As explained above,
this last detail is due to the fact the economic system is closed,
and then it conserves the total money, whose equivalent quantity in the geometrical
problem is $E$. If the economic system were open, with an upper limit in the wealth,
then the transformed system would evolve in an equiprobable way over the volume
defined by the inequality (\ref{eq-Ek}), although its statistical behavior
would continue to be the same as it has been proved for the cases $b=1,2$
in Refs. \cite{lopezruiz2008,lopezruiz2007}.

\section{Other additional geometrical questions}
\label{Sec4}

As two collateral results, we address two additional problems in this section.
The first one presents the finding of the general formula for the volume
of a high-dimensional symmetrical body and the second one offers an alternative
presentation of the canonical ensemble.

\subsection{Formula for the volume of a high-dimensional body}

We are concerned now with the asymptotic formula ($N\rightarrow\infty$) for the volume of
the $N$-dimensional symmetrical body enclosed by the surface
\begin{equation}
x_1^b+x_2^b+\cdots +x_{N-1}^b+x_N^b = E.
\label{eq-Ekk}
\end{equation}
The linear dimension $\rho$ of this volume, i.e., the length of one of its sides verifies
$\rho\sim E^{1/b}$.
As argued in Eq. (\ref{eq-volumenn}),
the $N$-dimensional volume, $V_N(b,\rho)$, is
proportional to the term $\rho^N$ and to a coefficient $g_b(N)$
that depends on $N$. Thus,
\begin{equation}
V_N(b,\rho)=g_b(N)\,\rho^N,
\label{eq-volumenn1}
\end{equation}
where the characteristic $b$ indicates the particular boundary given by equation (\ref{eq-Ekk}).

For instance, from Equation (\ref{eq-S_n1}), we can write in a formal way:
\begin{equation}
g_{b=1}(N) =  {1^{N\over 1}\over \Gamma({N\over 1}+1)}.
\label{eq-gN}
\end{equation}
From Eq. (\ref{eq-S_n3}), if we take the diameter, $\rho=2R$, as the linear dimension of
the $N$-sphere, we obtain:
\begin{equation}
g_{b=2}(N) =  {\left({\pi\over 4}\right)^{N\over 2}\over \Gamma\left({N\over 2}+1\right)}.
\label{eq-gN1}
\end{equation}
These expressions (\ref{eq-gN}) and (\ref{eq-gN1}) suggest a possible general formula
for the factor $g_{b}(N)$, let us say
\begin{equation}
g_{b}(N) =  {a^{{N\over b}}\over \Gamma\left({N\over b}+1\right)},
\label{eq-gN2}
\end{equation}
where $a$ is a $b$-dependent constant to be determined. For example,
$a=1$ for $b=1$ and $a=\pi/4$ for $b=2$.

\begin{figure}[t]
\caption{The factor $g_b(N)$ versus $b$ for $N=10, 40, 100$, calculated from Eq. (\ref{eq-gN5}).
Observe that $g_b(N)=0$ for $b=0$,
and $\lim_{b\rightarrow \infty}g_b(N)=1$.}
\centerline{\includegraphics[width=9cm]{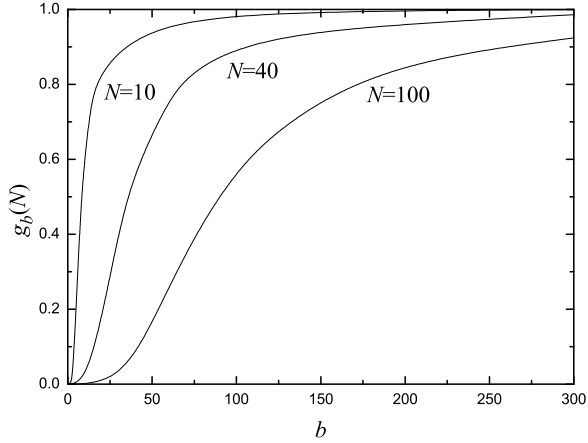}}
\label{fig2}
\end{figure}
In order to find the dependence of $a$ on the parameter $b$, the regime $N\rightarrow \infty$
is supposed. Applying Stirling approximation for the factorial $({N\over b})!$ in the denominator
of expression (\ref{eq-gN2}), and inserting it in expression (\ref{eq-cb}),
it is straightforward to find out the relationship:
\begin{equation}
c_b \,= \,(ab)^{-1/b}.
\label{eq-gN3}
\end{equation}
From here and formula (\ref{eq-cb2}), we get:
\begin{equation}
a\,= \,\left[\Gamma\left({1\over b}+1\right)\right]^b,
\label{eq-gN4}
\end{equation}
that recovers the exact results for $b=1,2$. The behavior of $a$
is monotonous decreasing when $b$ is varied from $b=0$,
where $a$ diverges as $a\sim 1/b+\cdots$,
up to the limit $b\rightarrow\infty$, where $a$ decays asymptotically
toward the value $a_{\infty}=e^{-\gamma}=0.5614$.

Hence, the formula for $g_{b}(N)$ is obtained:
\begin{equation}
g_{b}(N)\,= \,{\,\,\Gamma\left({1\over b}+1\right)^N\over \Gamma\left({N\over b}+1\right)},
\label{eq-gN5}
\end{equation}
It would be also possible to multiply this last expression (\ref{eq-gN5})
by a general polynomial $K(N)$ in the variable $N$,
and all the derivation done from Eq. (\ref{eq-gN2})
would continue to be correct. We omit this possibility
in our calculations. For a fixed $N$, we have that $g_b(N)$ increases monotonously
from $g_b(N)=0$, for $b=0$, up to $g_b(N)=1$,
in the limit $b\rightarrow\infty$ (see Fig. 2).
For a fixed $b$, we have that $g_b(N)$ decreases monotonously
from $g_b(N)=1$, for $N=1$, up to $g_b(N)=0$,
in the limit $N\rightarrow\infty$ (see Fig. 3).

The final result, that has been shown to be valid for any $N$ \cite{toral2009},
for the volume of an $N$-dimensional symmetrical body of
characteristic $b$ given by the boundary (\ref{eq-Ekk}) reads:
\begin{equation}
V_N(b,\rho)\,= \,{\,\,\Gamma\left({1\over b}+1\right)^N\over \Gamma\left({N\over b}+1\right)}\,\rho^N,
\label{eq-gN6}
\end{equation}
with $\rho\sim E^{1/b}$.

\begin{figure}[t]
\caption{The factor $g_b(N)$ versus $N$ for $b=10, 40, 100$, calculated from Eq. (\ref{eq-gN5}).
Observe that $g_b(N)=1$ for $N=1$,
and $\lim_{N\rightarrow \infty}g_b(N)=0$.}
\centerline{\includegraphics[width=9cm]{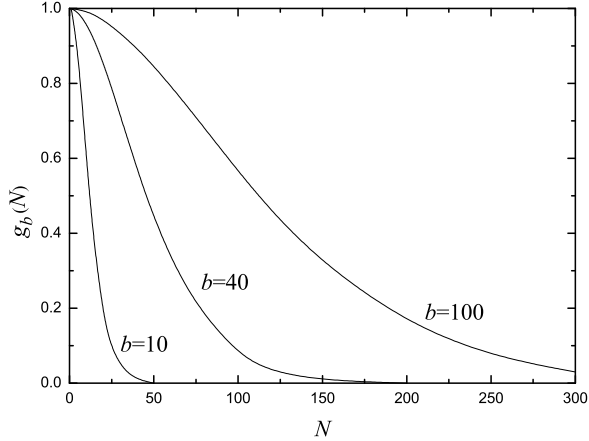}}
\label{fig3}
\end{figure}

\subsection{A microcanonical image of the canonical ensemble}

From Section \ref{Sec1}, here we offer a different image of the usual presentation
that can be found in the literature \cite{huang1987} about the canonical ensemble.

Let us suppose that a system with mean energy $\bar E$, and in thermal equilibrium
with a heat reservoir, is observed during a very long period $\tau$ of time.
Let $E_i$ be the energy of the system at time $i$. Then we have:
\begin{equation}
E_1+E_2+\cdots +E_{\tau-1}+E_{\tau} = \tau\cdot\bar E.
\label{eq-e4}
\end{equation}
If we repeat this process of observation a huge number (toward infinity) of times,
the different vectors of measurements, $(E_1,E_2,\ldots,E_{\tau-1},E_{\tau})$,
with $0\leq E_i\leq \tau\cdot\bar E$, will finish by covering equiprobably the
whole surface of the $\tau$-dimensional hyperplane given by Eq. (\ref{eq-e4}).
If it is now taken the limit $\tau\rightarrow\infty$, the asymptotic probability
$p(E)$ of finding the system with an energy $E$ (where the index $i$ has been removed),
\begin{equation}
p(E)\; \sim \;\; e^{-E/\bar E},
\label{eq-e22}
\end{equation}
is found by means of the geometrical arguments exposed in Section \ref{Sec1} \cite{lopezruiz2008}.
Doing a thermodynamic simile, the temperature $T$ can also be calculated. It is
obtained that
\begin{equation}
\bar E = kT.
\label{eq-e222}
\end{equation}
The {\it stamp} of the canonical ensemble, namely, the Boltzmann factor,
\begin{equation}
p(E)\;\sim\;\; e^{-E/kT},
\label{eq-e223}
\end{equation}
is finally recovered from this new image of the canonical ensemble.

\section{Conclusion}

In summary, this work has presented a straightforward geometrical argument
that in a certain way recalls us the equivalence between the canonical and
the microcanonical ensembles in the thermodynamic limit for the particular context
of physical sciences. In the more general context of homogeneous multi-agent systems,
we conclude by highlighting the statistical equivalence
of the volume-based and surface-based calculations in this type of systems.

Thus, we have shown that the Boltzmann factor or the Maxwellian distribution
describe the general statistical behavior of each small part
of a multi-component system in equilibrium whose components or parts are given
by a set of random linear or quadratic variables, respectively,
that satisfy an additive constraint,
in the form of a conservation law (closed systems) or
in the form of an upper limit (open systems), and that reach the
equiprobability when they decay to equilibrium.

Let us remark that these calculations do not need the knowledge of the exact or microscopic
randomization mechanisms of the multi-agent system in order to attain the equiprobability.
In some cases, it can be reached by random forces \cite{yakovenko2009}, in other cases
by chaotic \cite{bullard1992,pellicer2010} or deterministic \cite{gonzalez2008} causes.
Evidently, the proof that these mechanisms generate equiprobability is not a trivial task
and it remains as a typical challenge in this kind of problems.

The derivation of the equilibrium distribution for open systems in a general context
has also been presented by considering a general multi-agent system verifying an
additive constraint. Its statistical behavior has been derived from geometrical
arguments. Thus, the Maxwellian and the Boltzmann-Gibbs distributions are particular cases
of this type of systems. Also, other multi-agent economy models, such as
the Dragalescu and Yakovenko's model \cite{dragulescu2000},
the Chakraborti and Chakrabarti's model \cite{chakraborti2000} and
the modified Angle's model \cite{patriarca2006},
show similar statistical behaviors than our general geometrical system.
This fact has fostered our particular geometrical interpretation of all those models.

 We hope that this framework can be useful to establish other possible relationships
between the statistics of multi-agent systems and the geometry associated
to such systems in equilibrium.



\section*{Author details}

$^*$ L\'opez-Ruiz Ricardo \\
{\it Department of Computer Science, Faculty of Science,
Universidad de Zaragoza, Zaragoza, Spain\\
Also at BIFI, Institute for Biocomputation and Physics of Complex Systems,
Universidad de Zaragoza, Zaragoza, Spain}

Sa\~nudo Jaime\\
{\it Department of Physics, Faculty of Science,
Universidad de Extremadura, Badajoz, Spain\\
 Also at BIFI, Institute for Biocomputation and Physics of Complex Systems,
Universidad de Zaragoza, Zaragoza, Spain}


\end{document}